# Mining Data from the Congressional Record

Zhengyu Ma[1], Tianjiao Qi[1], James Route[1], Amir Ziai[1]

[1]School of Information, University of California, Berkeley, Berkeley, CA

tma@ischool.berkeley.edu, tqi@ischool.berkeley.edu, jroute@ischool.berkeley.edu, amir@ischool.berkeley.edu

**Abstract**

We propose a data storage and analysis method for using the US Congressional record as a policy analysis tool. We use Amazon Web Services and the Solr search engine to store and process Congressional record data from 1789 to the present, and then query Solr to find how frequently language related to tax increases and decreases appears. This frequency data is compared to six economic indicators. Our preliminary results indicate potential relationships between incidence of tax discussion and multiple indicators. We present our data storage and analysis procedures, as well as results from comparisons to all six indicators.

**Keywords:** Congress; tax; fiscal; policy; search; data; mining.

## 1. Introduction

US tax policy—whether for income, businesses, capital gains, or other forms—has long been a contentious issue in US politics. We often hear opposing viewpoints on taxes that are never reconciled, such as when to raise or lower taxes or whether tax adjustments should be aligned with business cycles to counteract recessions. Policy analyses, especially in the popular press, often focus on the timing of a single piece of legislation against the business cycle or an economic event [1]. We propose a method and architecture with a much broader scope. Instead of examining individual bills, we analyzed the entire Congressional record (as available) dating back to 1789 to investigate how legislative discussions on tax policy correlate with external factors, such as economic indicators and election results. Although this approach precludes drawing causal inferences, it can help identify patterns and relationships that warrant further study, such as how the onset of a recession or a change in party rule shapes US legislative discussions.

We are using data from GovTrack.us [2], a repository for bills and discussions in Congress dating back to the 18th century. We also use data on economic indicators, which we obtained from *Quandl* [3] in tabular format, and data on presidential election results from Wikipedia [4].

The majority of our storage and processing architecture runs on Amazon Web Services (AWS) to take advantage of its resiliency and flexible resources. We initially store all of our data in multiple Simple Storage Service (S3) buckets that function as a large, always-available repository. We run the Solr search engine on an Elastic Compute Cloud (EC2) instance, which indexes our data and enables us to perform positional text-based searches to locate discussions related to tax policy. We store the results of the searches locally and process them in R, along with our economic and political data, to conduct the final graphical and statistical analyses. Our preliminary results showed potential relationships between some of our economic indicators, such as S&P 500 returns and median household income. Section 4 presents our results, along with accompanying time series plots and a correlation matrix.

## 2. Acquisition and Organization of Data

We acquired approximately 21.5 GB of text-based data from GovTrack and used proximity searches to find terms of interest, such as "tax hike", and record their frequency of occurrence over time. The GovTrack data are stored in JSON and XML format. We used the rsync utility to copy the contents of GovTrack's directories, and then used the AWS sync utility to transfer the data to S3 buckets for later use in data processing. Although the GovTrack data are originally stored in a hierarchical file structure, the process of uploading to S3 converts them to a flat structure, making it simple to iterate through bucket contents and ingest data into Solr.

The GovTrack repository contains data on four different kinds of Congressional actions, each of which are listed below. Each individual file contains data associated with a single action. Each of these actions are documented in a differently-structured JSON or XML file.

- **Amendments**: Proposed changes to a bill presented before the House or Senate for action.

- **Bills**: Draft legislation that must pass a vote on the House and Senate floors in order to become law.

- **Nominations**: Candidates selected by the President for executive or judicial branch positions who must be approved by the Senate.

- **Votes**: Details of every measure submitted for a vote before the House and Senate.

We also collected data on six US economic indicators over the past several decades: annual GDP growth, median household income, unemployment rate, housing starts, S&P 500 returns, and the federal funds rate. We used an API from *Quandl* to collect the data using R and store the indicators as data frame objects. We also collected presidential election results from Wikipedia and stored the data in R with the economic indicators. We compared these data with the results of our proximity searches to determine if discussions of tax policy in Congress co-occurred with certain patterns of economic activity. Table 1 summarizes the data sources and retrieval tools that were used.

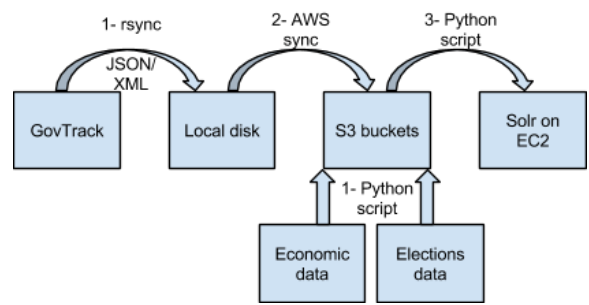

Figure 1. Data Acquisition and Storage Flowchart

Table 1. Data Source and Retrieval Tools

| Source | Description | Tool | Since | Size | Type |
|---|---|---|---|---|---|
| GovTrack | 1,000,000+ text-based files | rsync | 1789 | 21.5GB | JSON/XML |
| Quandl | 6 economic indicators | R | 1919 | < 100MB | R Dataframe |
| Wikipedia | Presidential election results | R | 1788 | 100KB | CSV |

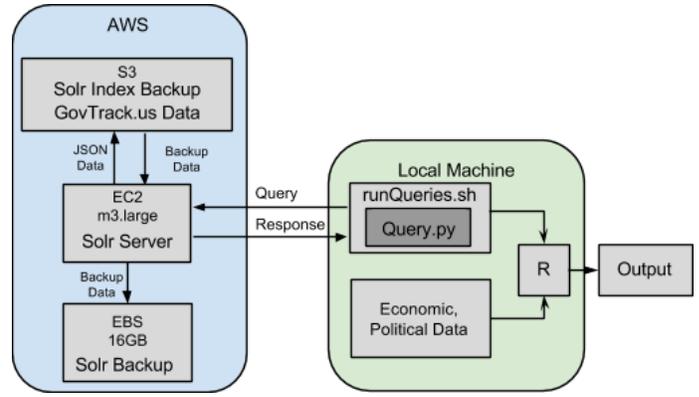

Figure 2. Architecture Overview

Figure 1 summarizes our acquisition and storage architecture. In Step 1, we acquire data from GovTrack to local disk and gather economic and elections data into S3 buckets. All of these actions in step 1 are performed in parallel. We then sequentially run steps 2 and 3 to upload GovTrack data from local disk into S3 buckets and then to create a search index in Solr.

2.1 Test Cases and Data Cleaning

We conducted multiple rounds of testing to ensure the integrity of the data. After downloading from GovTrack, we identified all of the data fields that appeared to be useful for indexing, and ensured that the data existed in all of the files. We used a Python script to crawl the JSON files to check that valid data could be extracted from all documents. There were many naming and formatting inconsistencies between the files, so we developed several logical cases to check for presence of various fields and process whatever data were available.

After validation, we began loading the GovTrack data into a Solr instance on EC2 for testing. We randomly sampled each data type (amendment, bill, nomination, and vote) from every Congressional session and ran positional searches in Solr to check that the sampled JSON document could be matched. This round of testing was successful and validated our approach for loading data into Solr.

3. Architecture Design

The majority of our architecture runs on AWS to take advantage of its resiliency and flexible computing and storage resources. However, some of the final, analytical tasks are performed locally in R to use specialized tools and record graphical outputs. Figure 2 provides an overview of our architecture. The data we stored in S3, as described in the previous section, serve as an input to the Solr server running on EC2. We back up the Solr data to both EBS and S3 so that the server can be quickly restored if a problem occurs. Once all data are ingested into Solr, we query the server from a local machine and store the results in CSV format for ease of processing. We then perform statistical comparisons using the economic and political data described in Section 2 to create our final outputs. The tools and libraries used throughout this project are listed in Table 2.

3.1 AWS Implementation Details

3.1.1 Solr Server

The majority of our processing occurred on an m3.large EC2 instance [5] running Solr; the remainder of the AWS architecture functions as static storage. This instance was sufficiently large to store the entire index of 500,000 JSON files in memory. We used the default configurations for the EC2 instance with additional whitelisting controls for security.

We ran Solr 5.0.0 as part of a pre-configured Bitnami stack [6]. Although Solr can run in a schema-less mode, we used Solr's basic schema to exercise more control over how Solr handles each of our data fields. We used Solr's dynamic fields rather than explicit fields, as these enabled us to control how Solr handles data from the client side, and made testing more efficient.

When inserting the JSON documents into Solr, we captured three primary chunks of information from each file. By indexing only the critical data from each file, we kept our index smaller and decreased the processing time required for indexing. The three data fields for each document are summarized below. Table 3 shows which specific fields from each type of JSON file are used to populate the fields in Solr.

Table 2. Tools and Libraries Used

| Stage | Tools / Libraries |
|---|---|
| Data Acquisition | rsync, R Quandl library |
| Data Storage | json, boto, S3, EBS |
| Data Processing and Indexing | Solr 5.0, EC2 Ubuntu instance, solrpy |
| Data Analysis | R, R ggplot2 library |

Table 3: Mapping of JSON Fields to Solr Index Fields

| Solr Field Types | JSON File Types | | | |
|---|---|---|---|---|
| | Amendment | Bill | Nomination | Vote |
| ID | amendment_id | bill_id | nomination Nomination | vote_id |
| Text | description purpose | description official_title | Nominee nominee Organization organization | question |
| Date | actions: acted_at | actions: acted_at | actions[1] | date |

- **id**: A string found within each JSON file that uniquely identifies it.
- **text_txt**: An array of strings containing all relevant text data from the document. The "_txt" suffix instructs Solr to treat the field as an array and apply stemming, stopword elimination, and positional indexing to the text in each element.
- **timestamp_dts**: An array of dates, expressed as strings, for each date an action associated with the document occurred in Congress. For example, an amendment could come up for discussion multiple times, causing multiple dates to be recorded for it (we want to account for each occurrence). The "_dts" suffix instructs Solr to treat the data as an array of specially formatted strings containing dates.

### 3.1.2 Solr Backup

We backed up the data in our EC2 instance and our S3 buckets to improve reliability in the event of a serious failure. We used a 16GB Elastic Block Store (EBS) drive connected to the EC2 instance, and we wrote the entire contents of the EC2 instance to EBS after each major data import to Solr (for example, after writing at least 50,000 files to the server). This made recovery simple if the EC2 instance went down or if we needed to transfer the configuration to another instance.

We also compressed the directory containing the Solr server's configuration files, index, and data stores, we and transferred the file to S3. This created more redundancy and was a simpler procedure than backing up the entire EC2 instance. It also enabled us to quickly transfer the index to another Solr server elsewhere if we needed to reproduce it quickly.

### 3.2 Local Processing Architecture

Our local architecture, shown in Figure 3, was simple and only used to perform the final stages of analysis (described in Section 4). We ran a local shell script, which was a wrapper around another Python script, to send search queries to the Solr server and obtain the data of interest from Solr. The Python script ran proximity searches on the contents of the text_txt field to find occurrences of phrases related to tax increases and decreases, and returned the ID strings and dates associated with every match. We removed duplicate results by examining the ID strings and aggregating by date; the final result was a CSV file that indicated the number of matches for each date.

We read this data into R to graph occurrences over time and overlay political and economic data, as described further in Section 4. We performed all of these tasks locally because the size of the CSV files were small (all less than 10MB) and the graphical environment for generating plots and statistical analyses was much easier to run locally than through an SSH connection to EC2.

### 3.3 Scalability

Our current architecture has plenty of room to grow if more data becomes available. The size of our individual data files is quite small (highly unlikely to hit S3's 5TB limit), so we can continue adding more files to buckets, or even creating new partitions, to store more data. Our EC2 instance also has plenty of room for growth. We only stored a portion of the data held in the GovTrack files, and approximately 80% of our system memory remained free. Therefore, we can accommodate at least a four-fold increase in data.

If the amount of data were to increase by a large amount (an order of magnitude or greater), we could simply run a more powerful EC2 instance to store and index it. We would have to increase the size of our EBS volume as well to hold backups. Even greater amounts of data would demand a different approach, such as using SolrCloud (see Section 5.1). In this case we would use multiple EC2 instances running Solr as a cluster to handle the data.

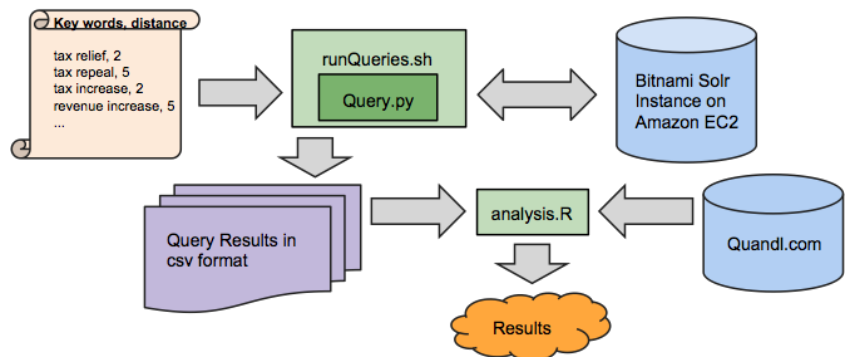

Figure 3. Local Architecture Overview

## 4. Analysis and Results

### 4.1 Querying the Solr Server

After Solr indexed all of the GovTrack data, we performed proximity searches. Our searches consisted of a pair of keywords and a term distance; we recorded the number of occurrences by date for each query and stored this data in CSV format. We explored a variety of tax-related terms and a range of distances. In order to pick the ones with the most meaningful and representative distributions over the years, we used shell commands (e.g. sed, awk) to do some simple statistics on the output files and compare their frequency distributions. The keyword pairs we focused on were: ("tax," "increase"), ("revenue," "increase"), ("tax," "relief"), and ("tax," "repeal"); the distances we used were 2-grams and 5-grams.

### 4.2 Data Aggregation

As summarized in Table 4, the datasets we collected all had different resolutions and availabilities. The tax term frequency data from our searches was aggregated at daily intervals and dated back to 1789. The economic indicators had daily, monthly, and annual resolutions, and the earliest one only dated back to 1919. The election results were available every four years and dated back to 1788. In order to graph their relationships over time and calculate meaningful correlation coefficients, we first aggregated each dataset into yearly intervals (our searches did not generate enough matches for meaningful comparisons at daily or monthly intervals). Then for each plot and correlation calculation, we used the earliest date when both time series were available as the starting date. As a result, the graphs for our economic indicators all had different starting dates. We show all [6 indicators] x [8 term frequency] = [48 plots] in Appendix A, and we included a few interesting graphs in the next section.

Table 4. Data Resolution and Availability

|  | **Resolution** | **Start Date** | **End Date** |
|---|---|---|---|
| Tax Term Frequencies | Daily | 1789-01-01 | 2015-03-31 |
| GDP Growth | Annual | 1961-12-31 | 2013-12-31 |
| Median Household Inc. | Annual | 1984-01-01 | 2013-01-01 |
| Unemployment Rate | Monthly | 1948-01-01 | 2015-03-01 |
| Housing Starts | Monthly | 1919-01-01 | 2015-03-01 |
| S&P 500 Index | Daily | 1950-01-03 | 2015-03-31 |
| Federal Fund Rate | Daily | 1954-07-01 | 2015-03-31 |
| Election Results | Every 4 years | 1788 | 2014 |

### 4.3 Summary of Results

Figure 4 presents four interesting time series graphs. The top left inset shows the relationship between median household income and the occurrence of "revenue" and "increase" within five words. We see a positive correlation between the two series, suggesting that when the household income increases, the Congress also tends to discuss increasing revenue. The shaded areas represent years when the Democratic Party controlled the presidency. (Overall, we did not observe any clear patterns between presidential party and tax policies.) The top right inset plot suggests that when the unemployment rate spiked, there were more political discussions on tax repeal. On the bottom left, we see that the S&P 500 index return seems to lead the tax repeal policies. The plot on the bottom right shows the relationship between federal funds rate and the occurrence of 'tax' and 'increase' within five words. We also observe a positive correlation, suggesting that when the federal funds rate is high, there was also more emphasis on tax relief.

Table 5 summarizes the correlation coefficients we calculated between the eight query types we ran and the six economic indicators we collected. Some interesting patterns arise. For example, median household income is positively correlated with both the ("revenue," "increase") 5-gram and ("tax," "relief") 5-gram, which suggests that when the household income rises, there are more debates over whether to raise or lower taxes. In addition, there is a consistent negative correlation between S&P 500 returns and tax decrease related terms. This could mean that when the financial markets are performing well Congress is less likely to discuss tax relief policies, or that changes in the financial markets tend to lead the discussions of tax cuts by a few years (as discussed earlier).

## 5. Improvements and Future Considerations

Our search architecture and analytical process appeared to perform well and yielded some interesting initial results. However, there remains room for improvement. If we were to conduct a follow-up study, there are a few areas that warrant further consideration:

- **Improve architecture resiliency**. Solr has additional features, such as SolrCloud, that can be enabled for sharding and resiliency.

- **Fill holes in the data**. GovTrack.us appears comprehensive, but the amount of data it holds is uneven from year to year.

- **Use more of the GovTrack data**. There are other fields in the dataset we acquired that we could use for further analysis.

We will examine each of these areas in detail, and where applicable discuss related work that we could use as an example for implementation.

### 5.1 Improving Resiliency

To better address resiliency, we could use SolrCloud [7], a feature that splits a collection across a computing cluster. Each node in the cluster runs its own Solr instance and holds a portion of the entire collection's index. With enough nodes in the cluster, SolrCloud not only implements sharding, but it replicates shards across multiple nodes. We would need to use multiple EC2 instances and manually install Solr rather than use the pre-configured Bitnami stack. Bitnami's implementation uses JAR files instead of a binary executable, which complicates the configuration process.

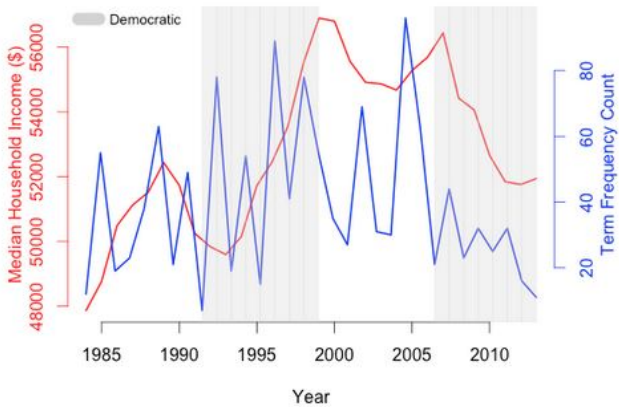
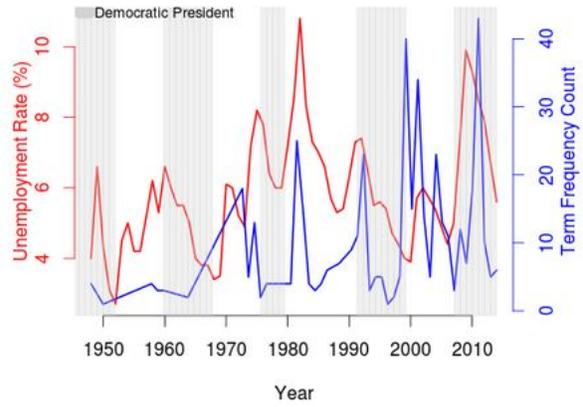
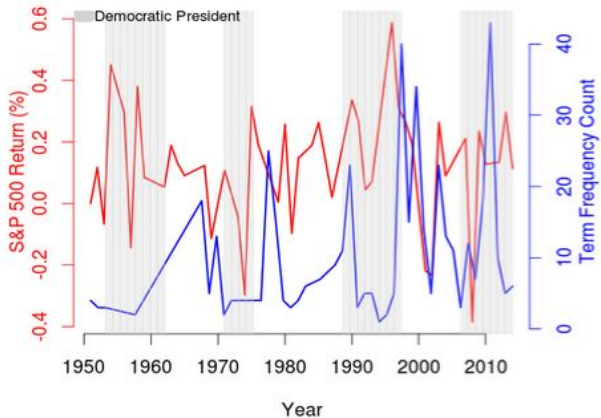
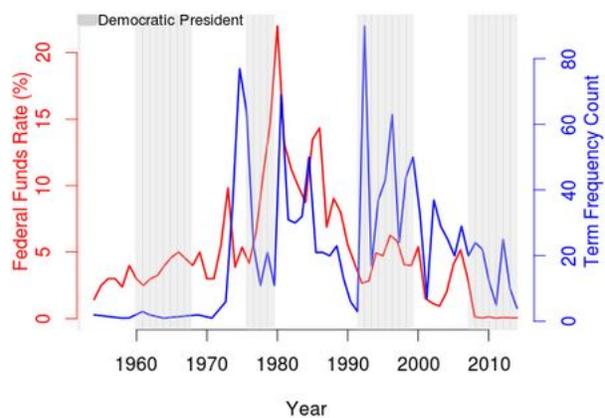

Figure 4 Sample Time Series Line Graphs

The benefit of SolrCloud is that if one replica goes down, the overall collection may still be able to function—as long the system acting as the ZooKeeper server, which organizes the nodes, is still working. However, this method is better suited for large datasets that face a much greater query load. In terms of practicality, our current method of backing up data is probably sufficient for the current scope of the project.

5.2  Filling Holes in the Data

The GovTrack data appears comprehensive and covers all sessions of Congress. However, its records from approximately 1970 onwards are far more detailed than in previous years, and from 1989 onwards there is even more data available. This biases the results, since the increase in total data means that we might see more matches from contemporary dates.

An example of a related project that faces similar difficulties is Capitol Words [8], which mines the Congressional Record to create visualizations of the most popular words spoken by lawmakers on the House and Senate floor. Capitol Words downloads the daily Congressional Record from the Government Printing Office (GPO). We could easily do the same by creating a web-scraping program and downloading the entire contents of the Record. Unfortunately, this is a partial solution because the GPO's online records only date back to 1994.

To obtain more historical material, we may have to use the National Archives and the Library of Congress as information sources. These institutions have additional records, such as the Annals of Congress and the Register of Debates, which are paraphrased or abridged forms of the Congressional Record that date back to the first sessions of Congress. Although not as comprehensive as the GPO, these sources are probably an improvement on what we currently have. These are not readily available online, and we would probably need to contact the curators for access. It is also unclear if electronic versions are available for all of these texts.

5.3  Expanding Use of the GovTrack Data

The GovTrack dataset contains additional information that we have not yet used, such as the voting record for all bills in both houses of Congress. We could conduct a social network analysis using the rolls calls and Congresspersons' voting records, as in [9], to determine the representatives who were the most influential and determine voting blocs. Taking this one step further, we could use the GovTrack data and identify long-term trends in voting outcomes for taxes (or any other issue, even), and then try to determine if there were certain lawmakers or groups who caused or influenced these trends. We would need to update our Solr index to include detailed voting results and roll calls in order to implement this feature.

Table 5. Correlation Between Term Frequency and Economic Indicators

| Search Term | GDP Growth | Median Household Income | Unemployment Rate | Housing Starts | S&P 500 Return | Federal Fund Rate |
|---|---|---|---|---|---|---|
| ("tax", "increase") 2-gram | -0.01 | 0.17 | 0.15 | 0.04 | 0.06 | 0.04 |
| ("tax", "increase") 5-gram | -0.03 | 0.04 | 0.19 | 0.08 | 0.11 | 0.15 |
| ("revenue", "increase") 2-gram | -0.03 | -0.08 | 0.22 | 0.22 | -0.19 | 0.11 |
| ("revenue", "increase") 5-gram | -0.21 | 0.38 | 0.12 | -0.06 | 0.00 | 0.17 |
| ("tax", "relief") 2-gram | -0.13 | 0.48 | -0.14 | 0.05 | -0.29 | -0.43 |
| ("tax", "relief") 5-gram | -0.14 | 0.53 | -0.15 | 0.01 | -0.33 | -0.44 |
| ("tax", "repeal") 2-gram | -0.23 | 0.24 | 0.14 | -0.02 | -0.17 | -0.19 |
| ("tax", "repeal") 5-gram | -0.06 | 0.33 | 0.22 | 0.11 | -0.09 | 0.01 |

Returning to the Capitol Words example, there are a few additional techniques we could try to expand our use of the GovTrack data. Capitol Words computes TF-IDF to determine the significance of individual words within its corpus. We could try this approach, and use the results to weight words and phrases in our data for importance. We probably would have to alter our techniques for storing data in Solr to accomplish this. Capitol Words splits its text corpus into n-grams (lengths between 1 and 5). We could re-process our data by splitting text into n-grams, calculating the TF-IDF for each, and adding the n-grams to Solr for indexing along with TF-IDF for weighting.

## 6. Conclusion

We proposed a method using a search engine and publically available Congressional and economic data to explore the relationship between U.S. tax policy and the country's economic situation. In order to efficiently handle the large volume of Congressional data, we used AWS for data storage and processing. We also took advantage of Solr's built-in positional indexing features to quickly mine the text data. We found some notable connections between the time series of tax discussion occurrences and several of the economic indicators. Our method is highly scalable and can readily accommodate an increase in data by at least a factor of four. By integrating additional sources of Congressional record data, we can improve on our preliminary findings.

## References

[1] M. Feldstein, "The effect of marginal tax rates on taxable income: a panel study of the 1986 Tax Reform Act", *Journal of Political Economy*, 1995.

[2] "GovTrack.us: Tracking the U.S. congress", *Retrieved from http://www.govtrack.us*, May 2015.

[3] "Quandl- find, use and share numerical data", *Retrieve from http://www.quandl.com*, May 2015.

[4] "United States presidential election - Wikipeia", *Retrieved from https://en.wikipedia.org/wiki/United_States_presidential_election*, May 2015.

[5] "AWS | Amazon EC2 | Instance Types", *Retrieved from http://aws.amazon.com/ec2/instance-types*, May 2015.

[6] Apache Solr Cloud Hosting, Apache Solr Hosting - Installers and VM", *Retrieved from http://bitnami.com/stack/solr*, May 2015

[7] "SolrCloud - Apache Solr reference guide - Apache software foundation", *Retrieved from http://cwiki.apache.org/confluence/display/solr/SolrCloud*, May 2015.

[8] "Capitolwords", Retrieved from http://capitolwords.org, May 2015.

[9] A. Jakulin, W. Buntine, T. M. La Pira and H. Brasher, "Analyzing the U.S. Senate in 2003: Similarities, Clusters, and Blocs", *Political Analysis*, Vol. 30, NO. 3, pp. 291-310, 2009.

# Appendix A. Time Series Plots of Tax-related Term Frequency vs Economic Indicators

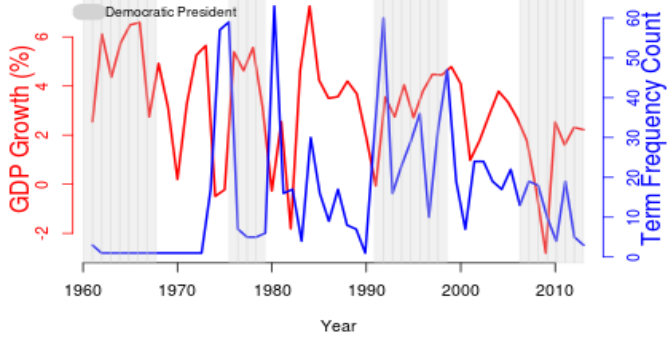
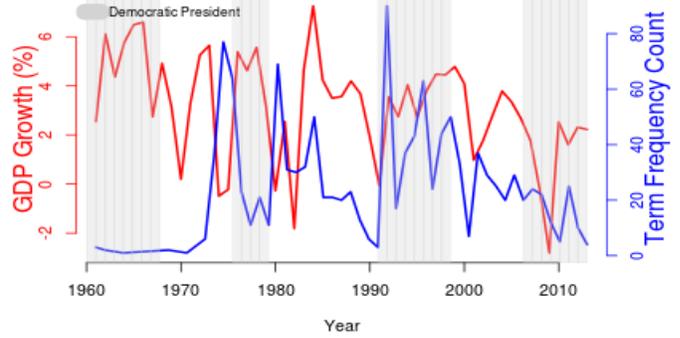
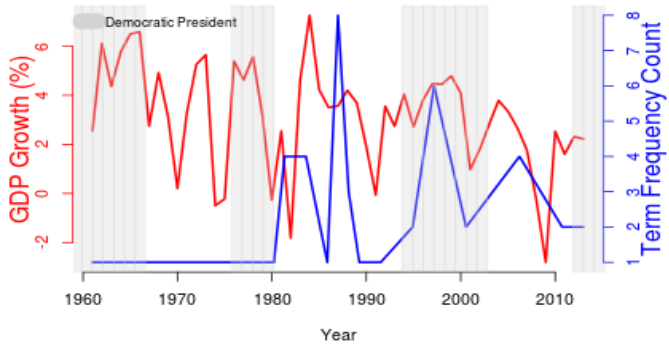
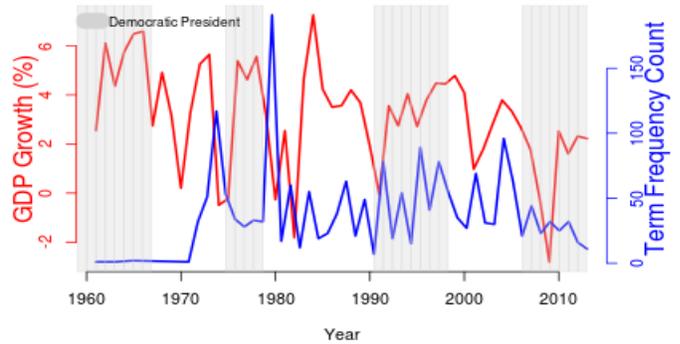
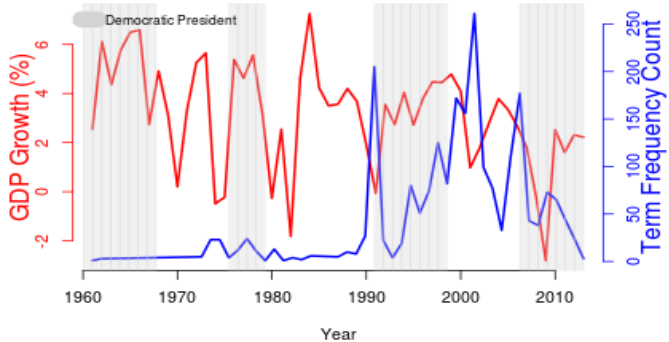
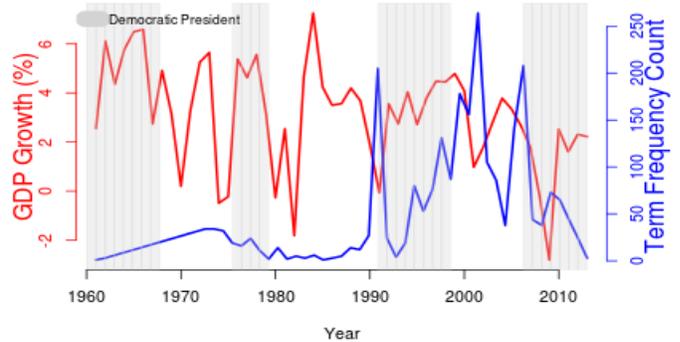
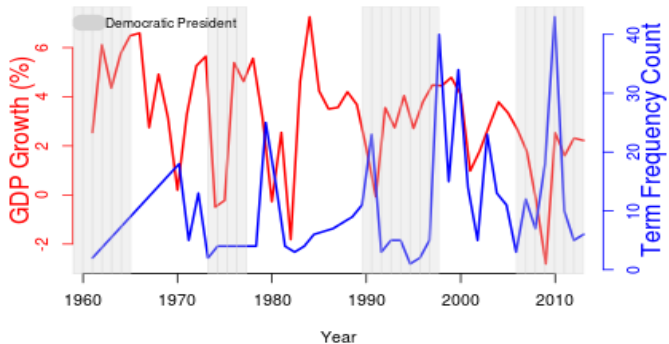
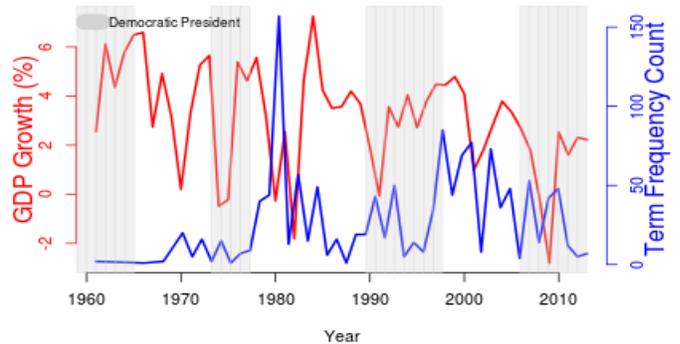

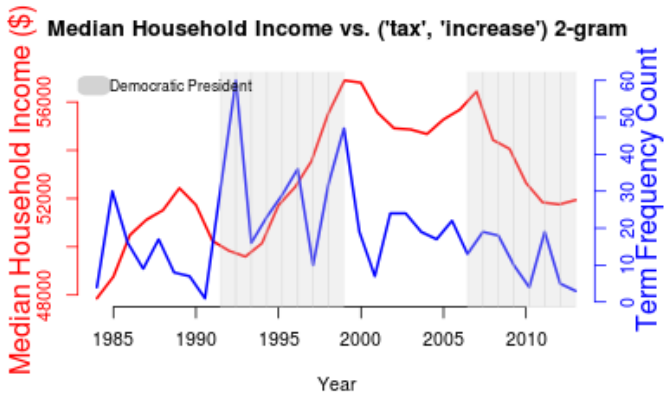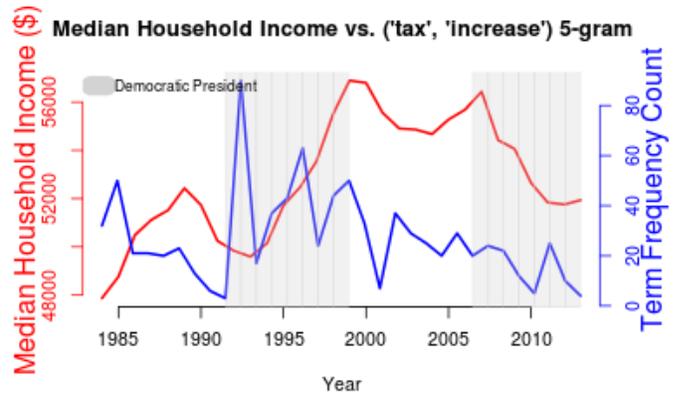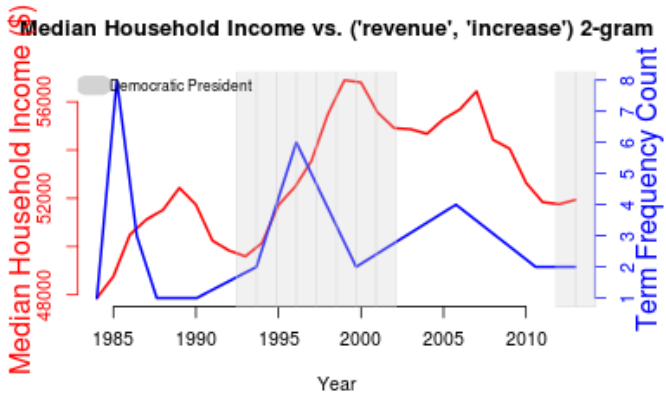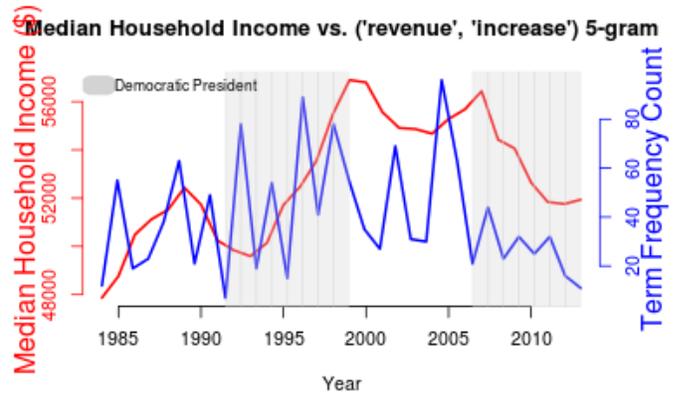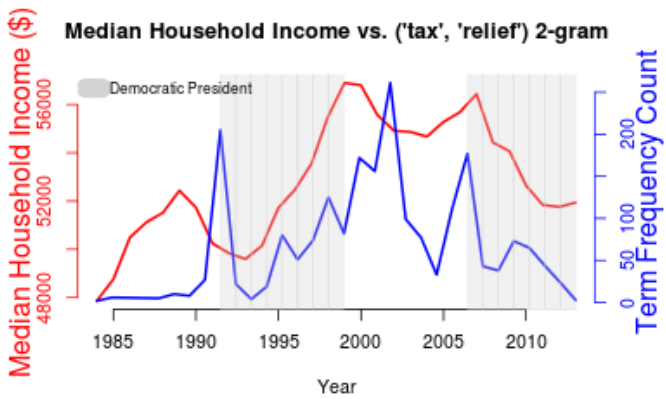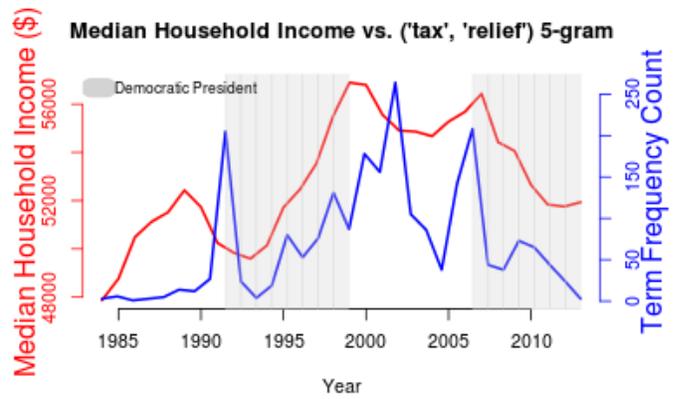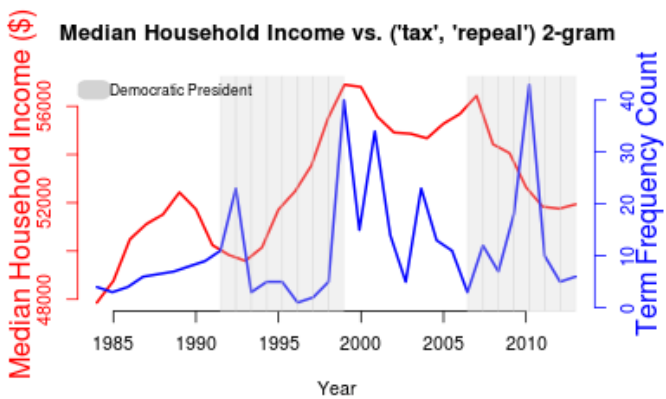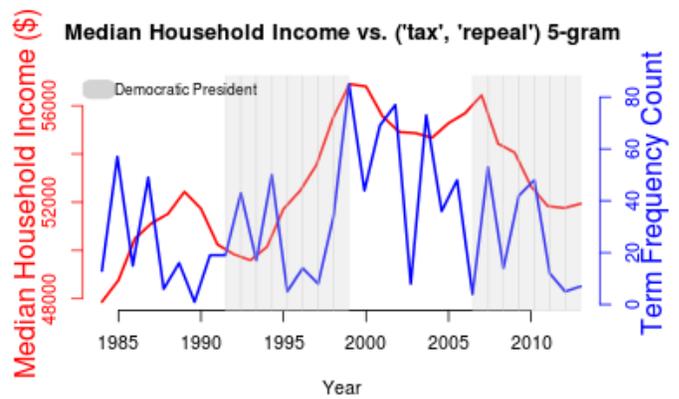

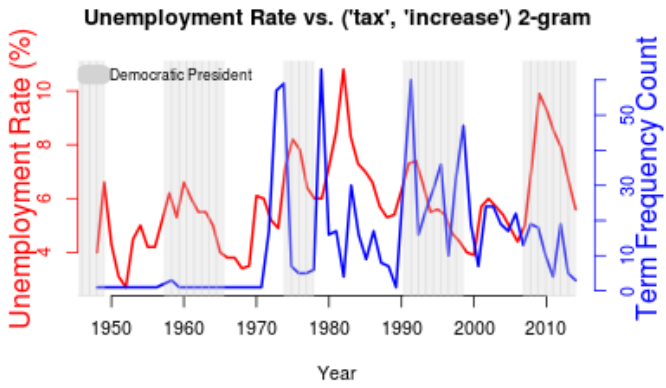
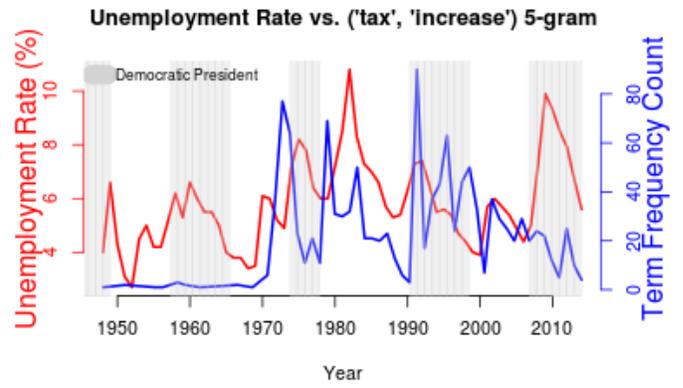
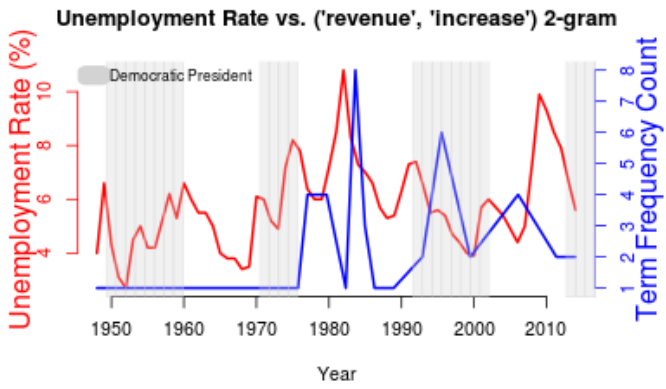
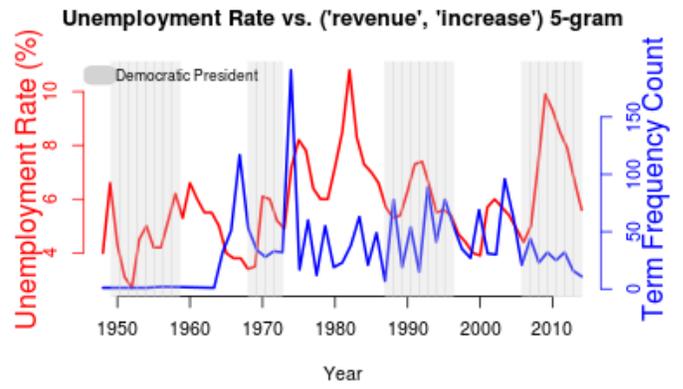
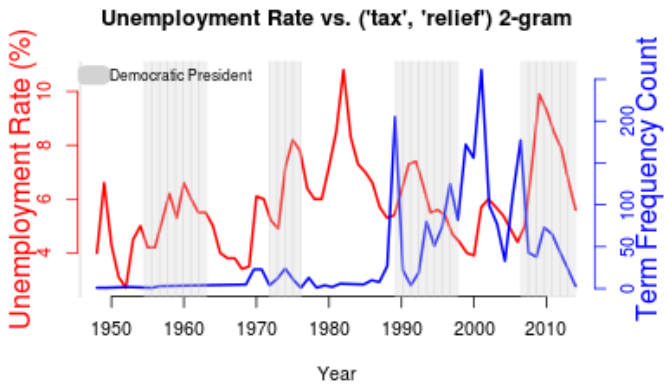
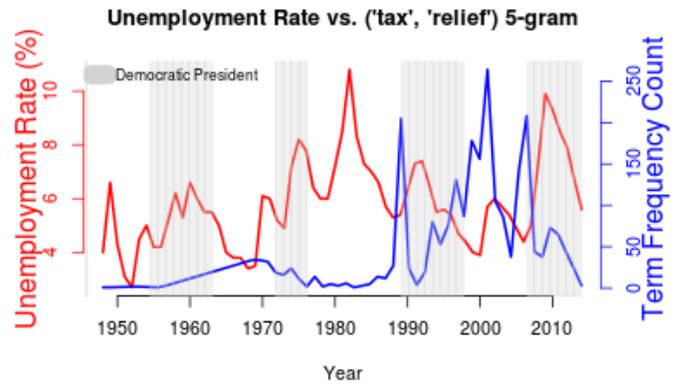
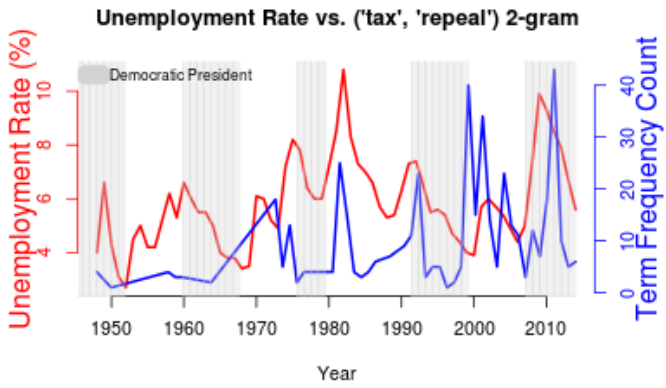
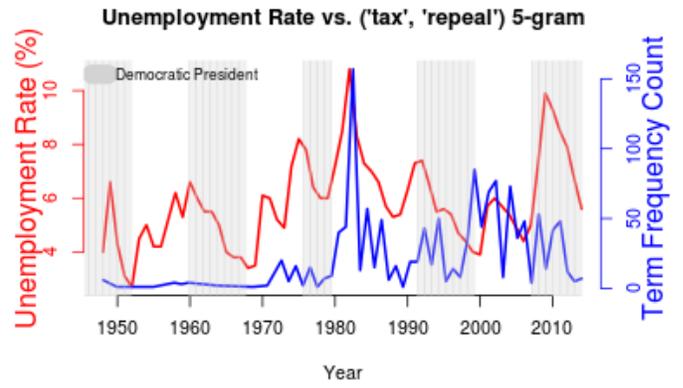

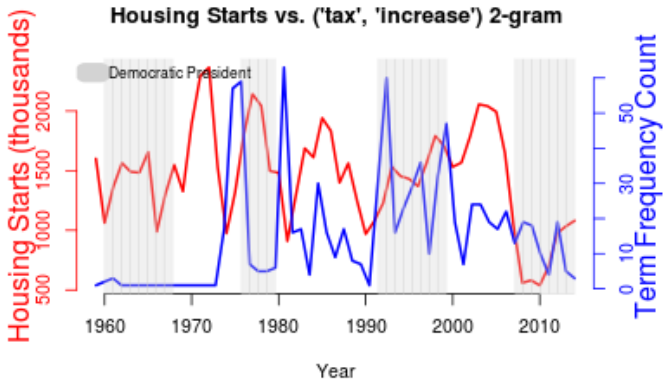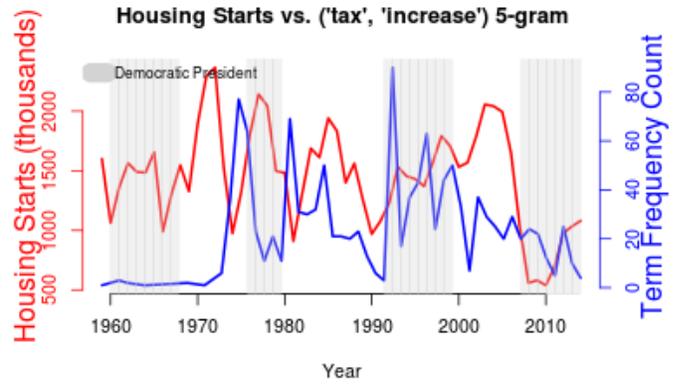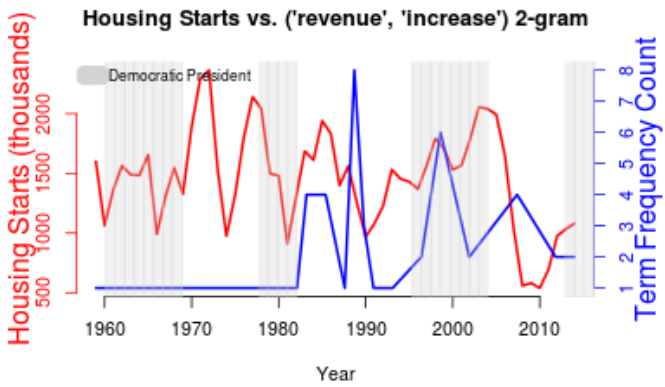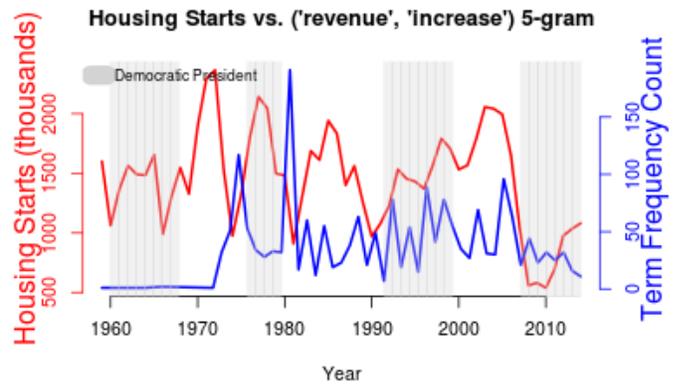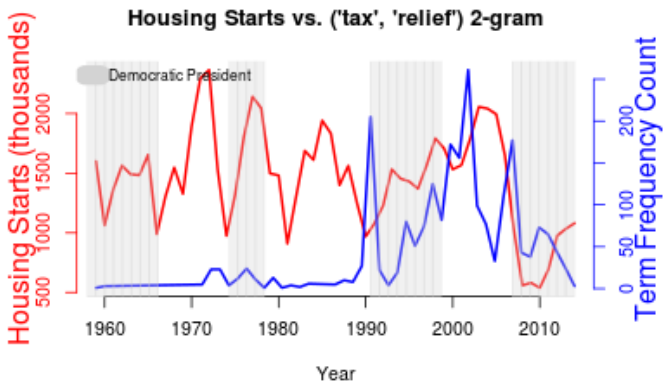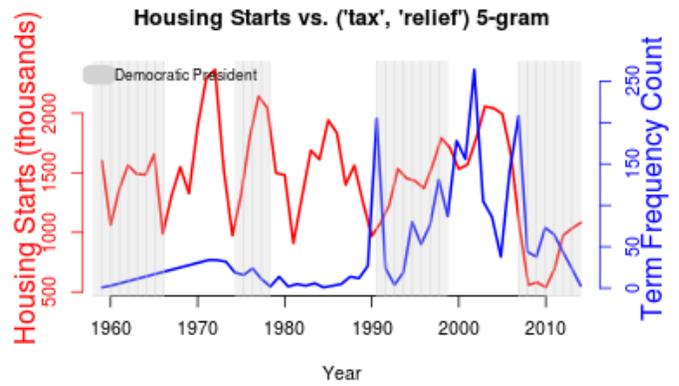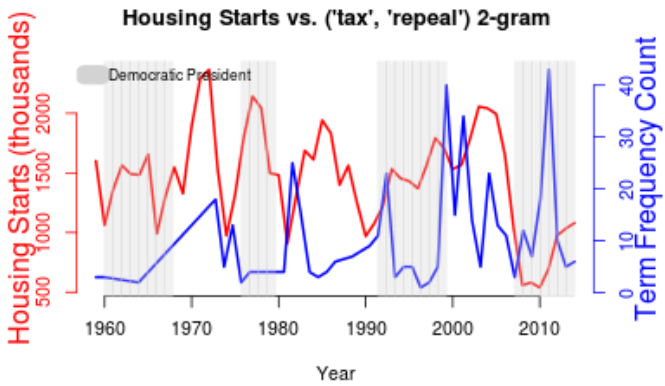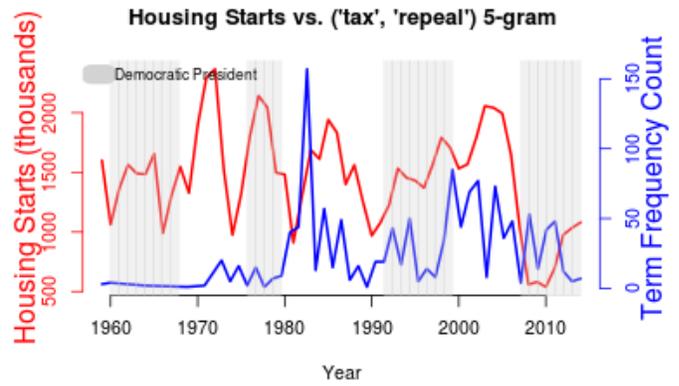

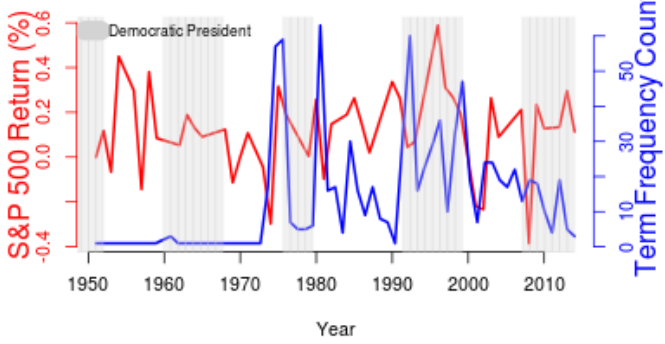
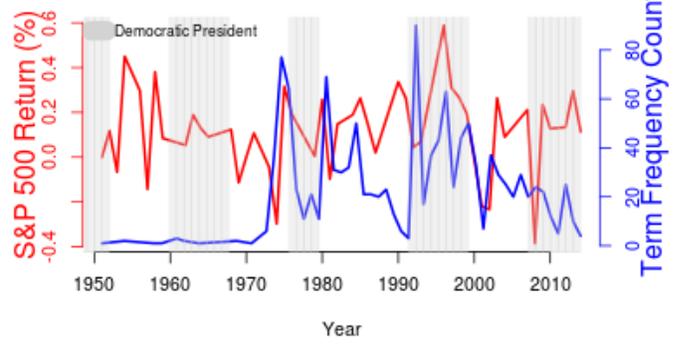
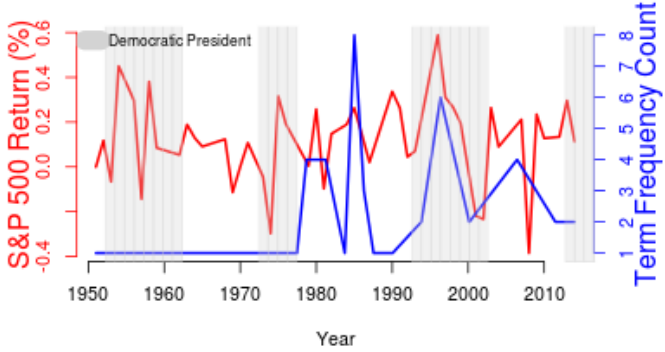
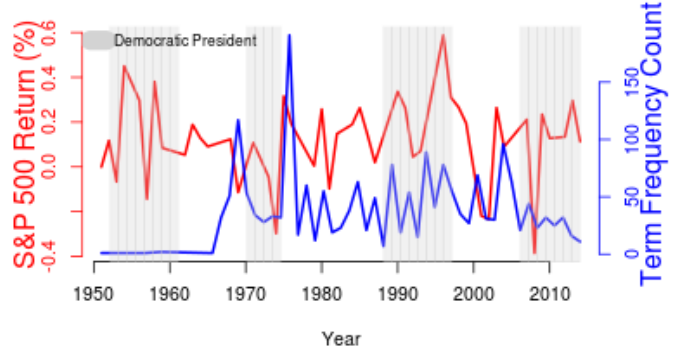
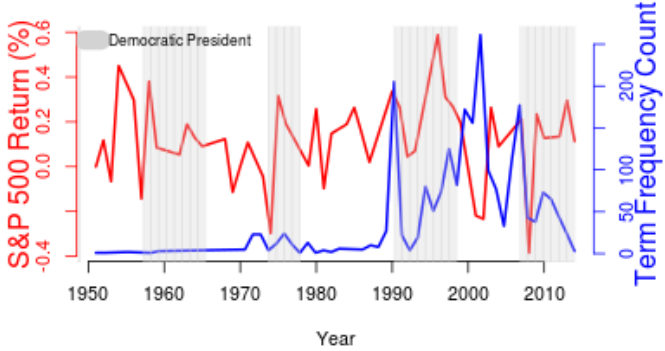
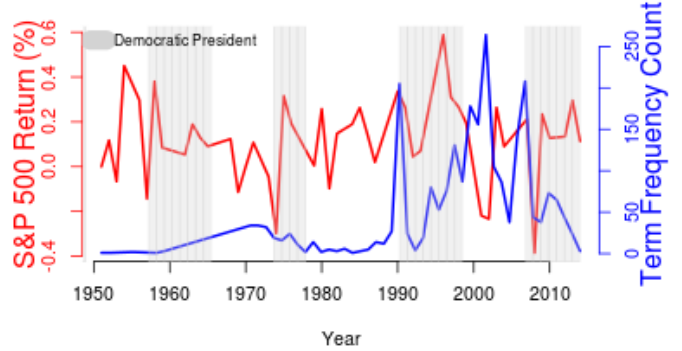
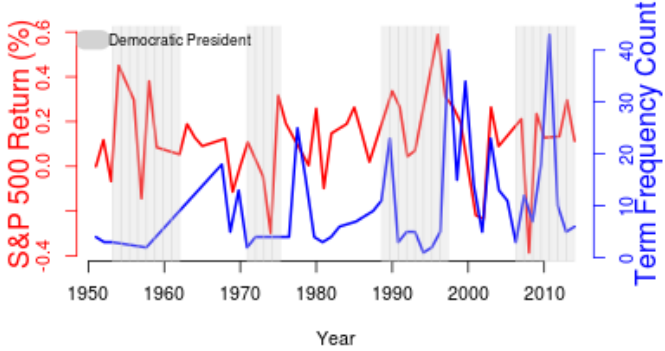
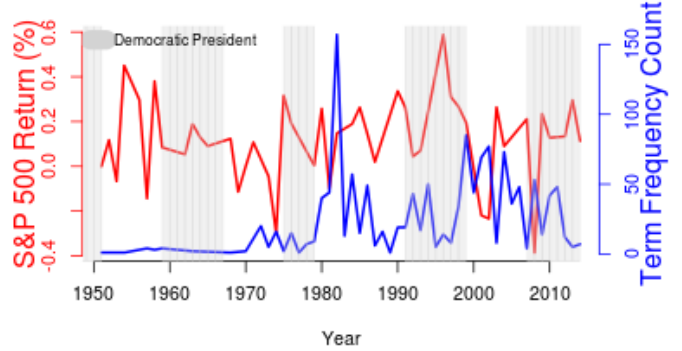

## Figures

**Federal Funds Rate vs. ('tax', 'increase') 2-gram**

**Federal Funds Rate vs. ('tax', 'increase') 5-gram**

**Federal Funds Rate vs. ('revenue', 'increase') 2-gram**

**Federal Funds Rate vs. ('revenue', 'increase') 5-gram**

**Federal Funds Rate vs. ('tax', 'relief') 2-gram**

**Federal Funds Rate vs. ('tax', 'relief') 5-gram**

**Federal Funds Rate vs. ('tax', 'repeal') 2-gram**

**Federal Funds Rate vs. ('tax', 'repeal') 5-gram**